\documentclass[10pt,letterpaper]{article}
\pdfoutput=1

\usepackage{jcappub}
\usepackage{amsmath,amssymb,color}
\usepackage{enumerate,xstring}
\usepackage{jcappub}
\usepackage{amsmath}
\usepackage{amsfonts}
\usepackage{amssymb}
\usepackage{graphicx}
\usepackage{epsfig}
\usepackage{color}
\usepackage{multirow}
\usepackage{graphicx}
\usepackage{bigints}
\usepackage{todonotes,bm,etoolbox}
\usepackage{calligra}
\usepackage{hyperref}
\usepackage{tabularx}
\usepackage{booktabs}
\usepackage{subfigure}

\usepackage{tikz}

\def\ba#1\ea{\begin{align}#1\end{align}}
\def\bea{\begin{eqnarray}}
\def\eea{\end{eqnarray}}
\def\be{\begin{equation}}
\def\ee{\end{equation}}

\def\({\left(}
\def\){\right)}
\def\[{\left[}
\def\]{\right]}

\def\<{\left\langle}
\def\>{\right\rangle}

\def\comment#1{}

\renewcommand{\v}[1]{\bm{#1}}

\def\vx{\v{x}}

\def\vq{{\v{q}}}

\newcommand{\perm}[1]{ \expandafter\ifstrempty\expandafter{#1} {\mbox{perm.}} {\mbox{$#1$ perm.}} }

\newcommand{\hi}{{\rm H_I}}
\newcommand{\fnl}{f_\textnormal{\textsc{nl}}}
\newcommand{\blin}{b_1}
\newcommand{\bphi}{b_\phi}
\newcommand{\bphidelta}{b_{\phi\delta}}
\newcommand{\A}{\mathcal{A}}

\definecolor{RedWine}{rgb}{0.743,0,0}
\definecolor{RoyalBlue}{rgb}{0.25,.41,.88}
\definecolor{ForestGreen}{rgb}{.13,.54,.13}
\definecolor{Goldenrod}{rgb}{.85,.65,.13}

\newcommand{\bq}{\begin{eqnarray}}
\newcommand{\eq}{\end{eqnarray}}

\title{\huge The local PNG bias of neutral Hydrogen, $\hi$}

\author[a,b]{Alexandre Barreira}

\affiliation[a]{\small Excellence Cluster ORIGINS, Boltzmannstra\ss e 2, 85748 Garching, Germany}
\affiliation[b]{\small Ludwig-Maximilians-Universit\"at, Schellingstra\ss e 4, 80799 M\"unchen, Germany}

\emailAdd{alex.barreira@origins-cluster.de}

\date{\today}

\abstract{We use separate universe simulations with the IllustrisTNG galaxy formation model to predict the local PNG bias parameters $\bphi$ and $\bphidelta$ of atomic neutral hydrogen, $\hi$. These parameters and their relation to the linear density bias parameter $b_1$ play a key role in observational constraints of the local PNG parameter $\fnl$ using the $\hi$ power spectrum and bispectrum. Our results show that the popular calculation based on the universality of the halo mass function overpredicts the $\bphi(b_1)$ and $\bphidelta(b_1)$ relations measured in the simulations. In particular, our results show that at $z \lesssim 1$ the $\hi$ power spectrum is more sensitive to $\fnl$ compared to previously thought ($\bphi$ is more negative), but is less sensitive at other epochs ($\bphi$ is less positive). We discuss how this can be explained by the competition of physical effects such as that large-scale gravitational potentials with local PNG (i) accelerate the conversion of hydrogen to heavy elements by star formation, (ii) enhance the effects of baryonic feedback that eject the gas to regions more exposed to ionizing radiation, and (iii) promote the formation of denser structures that shield the $\hi$ more efficiently. Our numerical results can be used to revise existing forecast studies on $\fnl$ using 21cm line-intensity mapping data. Despite this first step towards predictions for the local PNG bias parameters of $\hi$, we emphasize that more work is needed to assess their sensitivity on the assumed galaxy formation physics and $\hi$ modeling strategy.}

\begin{document}

\maketitle

\section{Introduction}
\label{sec:intro}

The level of non-Gaussianity of the distribution of the energy density fluctuations generated during inflation holds key information about the particle content and physics of the early Universe \cite{2004PhR...402..103B, 2010AdAst2010E..72C}. The most popular and well-studied type of this {\it primordial non-Gaussianity} (PNG) is called {\it local-type} PNG, where the primordial gravitational potential $\phi$ is written as \cite{2001PhRvD..63f3002K}
\bq\label{eq:fnl}
\phi = \phi_{\rm G} + \fnl\left[\phi_{\rm G}^2 - \left<\phi_{\rm G}^2\right>\right],
\eq
with $\phi_{\rm G}$ being a Gaussian random field and $\fnl$ a constant that quantifies the leading-order departure from Gaussianity ($\left<\cdots\right>$ indicates ensemble averaging). Different models of inflation make different predictions for $\fnl$, and so constraining its value observationally lets us distinguish between them. The current tightest bound comes from the analysis of the cosmic microwave background (CMB) data from the Planck satellite, which constrained $\fnl = -0.9 \pm 5.1\ (1\sigma)$ \cite{2020A&A...641A...9P}. Using large-scale structure data, the tightest constraint to date is $\fnl = -12 \pm 21\ (1\sigma)$ \cite{2021arXiv210613725M}, and was obtained using quasars from the eBOSS survey (see also Ref.~\cite{2019JCAP...09..010C}). This bound is still looser than the CMB one, but the volumes spanned by next-generation large-scale structure surveys should let us reach precisions of order $\sigma_{\fnl} \sim 1$, which would mark, even without a detection, an important landmark in our ability to distinguish between competing models of inflation \cite{2014arXiv1412.4671A, 2019Galax...7...71B}.

Some of these future constraints on $\fnl$ will come from studies of the spatial distribution of (atomic) neutral hydrogen ($\hi$) in the Universe \cite{2013PhRvL.111q1302C, 2015ApJ...812L..22F, 2015ApJ...814..145A, 2015PhRvD..92f3525A, 2018MNRAS.479.3490F, 2019PhRvD.100l3522B, 2020JCAP...11..052K, 2020MNRAS.496.4115C, 2021PhRvD.103f3520L, 2021PDU....3200821K, 2021arXiv210714057V}, which will be mapped by $21$cm line-intensity mapping experiments such as BINGO \cite{2021arXiv210701633A}, CHIME \cite{2014SPIE.9145E..22B}, HIRAX \cite{2016SPIE.9906E..5XN} and SKA \cite{2020PASA...37....7S}. Compared to traditional galaxy surveying where individual galaxies are identified on the sky and their redshifts are estimated spectroscopically or photometrically, the line-intensity mapping approach targets the integrated emission from all sources (resolved or not) in a given region of the Universe, with the radial information inferred through measurements in different frequency bands \cite{2017arXiv170909066K}. In the case of $\hi$, the relevant line is the $21$cm radiation emitted by its spin-flip transition. After the epoch of reionization $z \lesssim 6$, most of the hydrogen in the Universe is ionized, but a significant amount of $\hi$ is still present in galaxies inside gas clouds that are sufficiently dense to shield the $\hi$ from the ionizing radiation. By skipping the need to resolve individual galaxies, line-intensity mapping surveys can scan very rapidly large areas of the sky with good redshift coverage, which is key to probe the large distance scales where $\fnl$ contributes the most; when completed, some of these surveys will have mapped the $\hi$ out to redshifts $z\sim3$ over about half of the sky. These experiments are also subject to distinct observational systematics compared to traditional galaxy surveys (e.g.~the large foreground emission that needs to be subtracted), which will strongly establish the robustness of large-scale structure constraints on $\fnl$ if the different surveying techniques give consistent results \cite{2020MNRAS.499.4054C, 2021MNRAS.504..267F, 2021JCAP...04..081M}.

The key observational signatures of local PNG on the $\hi$ distribution arise through a set of {\it bias parameter} terms that are generated if $\fnl \neq 0$. In general, bias parameters are a set of redshift-dependent numbers that describe how the local abundance of a given tracer of the large-scale structure (dark matter halos, galaxies, $\hi$, etc.) is modulated by different types of long-wavelength ($\gtrsim 50-100\ {\rm Mpc}$) perturbations (see Ref.~\cite{biasreview} for a comprehensive review). For the case of $\hi$, and specifying already to the bias parameters we focus on in this paper, we can write \cite{dalal/etal:2008, mcdonald:2008, giannantonio/porciani:2010, assassi/baumann/schmidt}
\bq\label{eq:biasexp}
\rho_{\hi}(\vx, z)  \supset \bar{\rho}_{\hi}(z) \Big[1+  \blin(z)\delta_m(\vx, z) + \bphi(z)\fnl\phi(\vq) + \bphidelta(z)\fnl\phi(\vq)\delta_m(\vx, z) \Big],
\eq
where $\rho_\hi(\vx, z)$ is the local $\hi$ energy density, $\bar{\rho}_{\hi}(z)$ is its cosmic average, $\delta_m(\vx, z)$ is a large-scale total matter density fluctuation and $\phi(\vq)$ is a large-scale primordial gravitational potential fluctuation ($\vq$ is the initial Lagrangian coordinate associated with the final Eulerian coordinate $\vx$). This equation makes apparent the physical meaning of the bias parameters $\blin$, $\bphi$ and $\bphidelta$ as the {\it response} of the local $\hi$ energy density to the presence of the perturbation each multiplies, i.e., 
\bq\label{eq:respdef}
b_1 = \frac{1}{\rho_\hi} \frac{\partial\rho_\hi}{\partial\delta_m}\ \ \ \ ; \ \ \ \ \bphi = \frac{1}{\rho_\hi}\frac{\partial\rho_\hi}{\partial(\fnl\phi)}\ \ \ \ ; \ \ \ \ \bphidelta = \frac{1}{\rho_\hi} \frac{\partial^2\rho_\hi}{\partial(\fnl\phi)\partial\delta_m}.
\eq
For example, $b_1$ quantifies how much more ($b_1 > 0$) or less ($b_1 < 0$) $\hi$ exists inside large-scale mass overdensities. These three parameters are the most important ones for observational constraints on $\fnl$ using the $\hi$ power spectrum ($2$-point function) and bispectrum ($3$-point function). Concretely, for the case of the $\hi$ power spectrum, and analogously to the case of the galaxy power spectrum \cite{dalal/etal:2008}, the $\fnl$ signature is $\propto b_1\bphi\fnl/k^2$, where $k$ is the wavenumber.\footnote{This is often referred to as the {\it scale-dependent} bias effect, which can be a misleading name since what is scale-dependent is not any bias parameter, but the relation between potential and mass perturbations $\phi \sim \delta_m/k^2$.} Similarly, the leading-order contributions to the bispectrum include a series of terms proportional to $\fnl\blin^3$, $\fnl\bphi\blin^2$ and $\fnl\bphidelta\blin^2$, whose amplitude is what can be used to constrain $\fnl$ (see e.g.~Fig.~1 of Ref.~\cite{2020JCAP...12..031B} for a visualization of these contributions).

Since the observational signatures of $\fnl$ are effectively degenerate with the bias parameters, a solid understanding of the latter is critical to obtain the best possible constraints on local PNG from large-scale structure data (see Refs.~\cite{2020JCAP...12..031B, 2021JCAP...05..015M, 2022JCAP...01..033B} for recent discussions). However, as the bias parameters effectively describe how the long-wavelength environment impacts the $\hi$ distribution, they depend on the complicated interplay between gravity, reionization, star formation and stellar/black hole feedback, and are thus extremely challenging to predict theoretically. While $\blin$ can be fitted for directly from the data since it enters also through terms that do not multiply $\fnl$, the same is not true for the local PNG parameters $\bphi$ and $\bphidelta$, which require assumptions to be made on in order to constrain $\fnl$ competitively. The standard way around this problem is based on the hope that, while the parameters $\bphi$ and $\bphidelta$ themselves are expected to be extremely sensitive to the details of structure formation, their relation to $\blin$ may be more robust and easier to predict theoretically. Thus, if $\bphi$ and $\bphidelta$ can be fixed in terms of $b_1$, then determining the latter from the data breaks the degeneracy with $\fnl$, which can then be constrained. 

For example, the majority of the constraint studies on $\fnl$ in the literature (including those based on $\hi$ data) assume relations based on the universality of the halo mass function, from which it follows e.g.~that $\bphi \propto \left(b_1 - 1\right)$. Despite its widespread adoption however, there is no reason to expect these relations to hold exactly for real tracers of the large-scale structure like the $\hi$, which motivates simulation-based works to refine these simple theoretical priors. Indeed, even for the case of dark matter halos in gravity-only simulations, we now know that $\bphi(b_1)$ \cite{grossi/etal:2009, 2017MNRAS.468.3277B} and $\bphidelta(b_1)$ \cite{2022JCAP...01..033B} are not perfectly described by the corresponding universality expressions. Further, using hydrodynamical simulations of galaxy formation, Refs.~\cite{2020JCAP...12..013B, 2022JCAP...01..033B} showed recently that the $\bphi(b_1)$ and $\bphidelta(b_1)$ relations of galaxies selected by a variety of criteria (including stellar-mass, black-hole mass accretion rate and color) are also not adequately described by the universality assumption, which needs to be revisited.

This motivates our main goal in this paper, which is to use hydrodynamical simulations to study the bias parameters $\bphi$ and $\bphidelta$ of the $\hi$ distribution. The methodology we adopt in this paper is similar to that of Refs.~\cite{2020JCAP...12..013B, 2022JCAP...01..033B}, who used separate universe simulations of galaxy formation with the IllustrisTNG model to study the same bias parameters, but for the halo and galaxy distribution. One of our main results is that the $\bphi(b_1)$ and $\bphidelta(b_1)$ relations of the $\hi$ are generically overpredicted by the results that are typically encountered in the literature assuming the above-mentioned universality expressions. We show this can be explained physically by the impact that long-wavelength $\fnl\phi$ perturbations have on the $\hi$ content of dark matter halos, which can be understood as a new {\it bias} parameter (or response function) in the halo model. We will show results for just a single galaxy formation model and $\hi$ modeling recipe, and as a result, given the uncertainties we still have on these areas, and despite the intuition this will let us build already, the analysis in this paper should be regarded as the first step towards robust theoretical priors on the $\bphi(b_1)$ and $\bphidelta(b_1)$ relations of the $\hi$. We stress that these relations play a key role in observational constraint studies of $\fnl$ using $\hi$ data, which strongly motivates more works like this. The rest of this paper is organized as follows. In Sec.~\ref{sec:method}, we describe the galaxy formation simulations, the $\hi$ modeling strategy, and the numerical methods that we use to estimate the $\hi$ bias parameters $b_1$, $\bphi$ and $\bphidelta$. We show and discuss our numerical results in Sec.~\ref{sec:results}, and summarize and conclude in Sec~\ref{sec:conc}.

\section{Methodology}
\label{sec:method}

In this section we describe the $N$-body simulations, the modeling of the $\hi$ distribution and the methods we use to estimate the $\hi$ bias parameters $\blin$, $\bphi$ and $\bphidelta$ defined in Eqs.~(\ref{eq:biasexp}) and (\ref{eq:respdef}).

\subsection{Numerical simulation specifications}
\label{sec:sims}

We obtain our numerical results using separate universe simulations run with the moving-mesh code {\sc AREPO} \citep{2010MNRAS.401..791S} and IllustrisTNG as the galaxy formation model \citep{2017MNRAS.465.3291W, Pillepich:2017jle}. This model is characterized by a set of {\it sub-grid} prescriptions for physics including gas cooling, star formation, metal enrichment, stellar winds, supernovae feedback, and black hole feedback, that broadly reproduces a series of key observations including the low-redshift galaxy stellar mass function, the cosmic star formation history, the sizes and colors of galaxies and the gas fractions in galaxies and galaxy groups \cite{2018MNRAS.480.5113M, Pillepich:2017fcc, 2018MNRAS.477.1206N, 2018MNRAS.475..676S, Nelson:2017cxy, 2019MNRAS.490.3234N, 2019MNRAS.490.3196P}. The fiducial cosmological parameters are: mean total matter density today $\Omega_{m0} = 0.3089$, mean baryon density today $\Omega_{b0} = 0.0486$, mean dark energy density today $\Omega_{\Lambda0} = 0.6911$, dimensionless Hubble rate $h = 0.6774$, primordial scalar power spectrum amplitude $\A_s = 2.068 \times 10^{-9}$ (at $k_{\rm pivot} = 0.05/{\rm Mpc}$, corresponding to $\sigma_8(z=0) = 0.816$), and primordial scalar spectral index $n_s = 0.967$.

As we describe below, we measure the $\hi$ bias parameters $\bphi$ and $\bphidelta$, respectively, as the response of the $\hi$ abundance and its linear bias parameter $\blin$ to long-wavelength $\fnl\phi$ perturbations. Under the separate universe approach, this is equivalent to the response to changes in the cosmological parameter $\A_s$, and so we consider also two additional cosmologies called High$\A_s$ and Low$\A_s$, which share the same parameters as the fiducial except that  $\A_s \to \A_s\left[1 + \delta_{\A_s}\right]$, where $\delta_{\A_s} = +0.05$ for High$\A_s$ and $\delta_{\A_s} = -0.05$ for Low$\A_s$. When we vary $\A_s$, we keep the sub-grid parameters of the IllustrisTNG model fixed also to their fiducial values in order to interpret our results on the impact of $\fnl\phi$ perturbations at fixed galaxy formation physics.

We will show results obtained at two numerical resolutions. One is called TNG300-2 and is characterized by a cubic box with size $L_{\rm box} = 205\ {\rm Mpc}/h$ and $N_{\rm p} = 2\times 1250^3$ initial dark matter and gas resolution elements. The other is higher-resolution, it is called TNG100-1.5, and it is characterized by $L_{\rm box} = 75\ {\rm Mpc}/h$ and $N_{\rm p} = 2\times 1250^3$. These separate universe simulations have been presented for the first time in Ref.~\cite{2020JCAP...12..013B} (to which we refer the reader for more details), and used in previous works already to study galaxy bias \cite{2022JCAP...01..033B} and halo occupation distribution responses \cite{Voivodic:2020bec}. In our results below we analyse the simulation outputs at redshifts $z = 0, 0.5, 1, 2, 3$. We have also used the publicly available IllustrisTNG data for the resolutions dubbed TNG300-1 ($L_{\rm box} = 205\ {\rm Mpc}/h$, $N_{\rm p} = 2\times 2500^3$) and TNG100-1 ($L_{\rm box} = 75\ {\rm Mpc}/h$, $N_{\rm p} = 2\times 1820^3$) to carry out some numerical convergence and consistency checks of our modeling at the fiducial cosmology.

When we analyse the $\hi$ content in halos, we consider gravitationally-bound objects found using a friends-of-friends algorithm with linking length $b = 0.2$ times the mean dark matter interparticle distance. We quote as halo mass $M_{\rm h}$ and total $\hi$ mass $M_{\hi}$ the summed contribution from all halo member particles and cells, and consider objects that comprise at least $N_{\rm cell} \geq 50$ gas cells.

\subsection{Modeling of $\hi$}
\label{sec:hi}

Our strategy to model the $\hi$ distribution in our simulations follows largely that of Ref.~\cite{2018ApJ...866..135V}. In IllustrisTNG, the hydrogen gas mass fraction of each Voronoi cell starts off at its primordial value of $f_{\rm H} = 0.76$, and it decreases with time as stars form and enrich the interstellar medium with heavier metals. The first part of the modeling of $\hi$ involves determining the fraction $f_{\rm H_N}$ of all hydrogen that is neutral. For non-star-forming gas (which for the star formation model in IllustrisTNG \cite{2003MNRAS.339..289S} is all gas with local hydrogen number density $n_{\rm H} < 0.106/{\rm cm}^3$),  we use the split into neutral and ionized fractions that is calculated self-consistently during the simulation taking into account the redshift-dependent UV radiation background, attenuation due to self-shielding in higher-density gas regions, and ionizing radiation by local active galactic nuclei (AGN) \cite{2013MNRAS.436.3031V}.

The situation is more complicated for star-forming gas because $f_{\rm H_N}$ is not computed self-consistently: the simulation output values are based on a mass-weighted temperature of the cold and hot gas phases, which is not a well-defined physical temperature. The authors of Ref.~\cite{2018ApJ...866..135V} go about this complication by setting the temperature of all star-forming gas cells to $T = 10^4 {\rm K}$, and repeating the default calculation $f_{\rm H_N}$ at post-processing with this temperature value. Here, based on the discussion in Ref.~\cite{2018ApJS..238...33D}, we follow an even simpler approach and consider {\it all hydrogen of star forming cells to be neutral}. This is based on the observation that the hot gas in star-forming regions is expected to be entirely ionized, the cold component is mostly neutral, and that the star formation model of Ref.~\cite{2003MNRAS.339..289S} predicts this cold fraction to be within 0.9 and 1 (cf.~the top right panel of Fig.~1 in Ref.~\cite{2018ApJS..238...33D} for an illustration of how good an approximation $f_{\rm H_N}=1$ is for star-forming gas).

Given the neutral hydrogren fraction, the second part of the modeling involves determining the fraction of ${\rm H_N}$ that is in atomic $\hi$ and molecular ${\rm H}_2$ form. As in Ref.~\cite{2018ApJ...866..135V}, we use the Krumholz-McKee-Tumlinson (KMT) model \cite{2008ApJ...689..865K, 2009ApJ...693..216K, 2010ApJ...709..308M} to determine the fraction $f_{\rm H_2}$ of all neutral Hydrogen that is molecular:
\bq\label{eq:kmt1}
f_{\rm H_2}=\begin{cases}
          1 - {0.75s}/({1 + 0.25s}) \quad &\text{if} \, s < 2 \\
          0 \quad &\text{if} \, s \geq 2\ \text{or\ gas\ is\ non-star-forming}, \\
     \end{cases}
\eq
with
\bq\label{eq:kmt2}
s &=& \frac{{\rm ln}\left(1 + 0.6\chi + 0.01\chi^2\right)}{0.6\tau_c}, \\
\chi &=& 0.756\left(1 + 3.1Z^{0.365}\right), \\
\tau_c &=& \frac{3}{4}\frac{\Sigma_{\rm gas} \sigma_d}{\mu_{\rm H}},
\eq
and where $Z$ is the gas metalicity in solar units, $\sigma_d = Z\times 10^{-21}\ {\rm cm}^2$ is an estimate of the cross-section of dust, $\mu_{\rm H}$ is the mean hydrogen nucleus mass and $\Sigma_{\rm gas}$ is the gas surface density. The latter can be calculated as $\Sigma_{\rm gas} = \lambda_{\rm J}\rho_{\rm gas}$, where $\lambda_{\rm J}$ is the gas Jeans length (used here as a proxy for the size of gravitationally-bound gas clouds) and $\rho_{\rm gas}$ is the gas volume density, or using a more sophiscated approach based on actually projecting the three-dimensional gas distribution in the simulation \cite{2018ApJS..238...33D, 2019MNRAS.487.1529D}. Here, we follow the approximate approach of Ref.~\cite{2018ApJ...866..135V} and evaluate the gas surface density as $\Sigma_{\rm gas} = R\rho_{\rm gas}$, with $R = \left(3V_{\rm cell}/(4\pi)\right)^{1/3}$ and $V_{\rm cell}$ the volume of the Voronoi gas cell in {\sc Arepo}. Finally, the atomic neutral hydrogen fraction we are interested in is given by $f_{\hi} = 1 - f_{\rm H_2}$, and we compute the $\hi$ energy density as $\rho_{\hi} = f_{\hi}f_{\rm H_N}\rho_{\rm H}$, where $\rho_{\rm H}$ is the total hydrogen density in each gas cell.

It should be noted that the KMT model is just one of several different approaches to model the $\hi$ and ${\rm H_2}$ phases of neutral hydrogen in hydrodynamical simulations. More advanced analytical approaches can account in particular also for the contribution of young stars to the ionizing radiation field in the interstellar medium, and other approaches based on empirical observational correlations and calibration against high-resolution numerical simulations also exist; see Refs.~\cite{2018ApJS..238...33D, 2019MNRAS.487.1529D} for a comparison of such different models using also IllustrisTNG. The simple KMT model can nonetheless be regarded as representative enough of the phenomenology of all these different approaches, and sufficient to our purpose here to begin to build intuition about the local PNG bias parameters of the $\hi$. In fact, Fig.~13 of Ref.~\cite{2018ApJS..238...33D} shows that the dependence of $f_{\rm H_2}$ on the local hydrogen density predicted by the KMT model is in between the predictions from many other models, which Ref.~\cite{2019MNRAS.487.1529D} subsequently showed perform roughly in the same way when compared to the data. This justifies the adoption of the KMT model in this paper, but we note that future analyses of these $\hi$ bias parameters should eventually determine the impact of different $\hi$ modeling techniques.

\begin{figure}
\centering
\includegraphics[width=\textwidth]{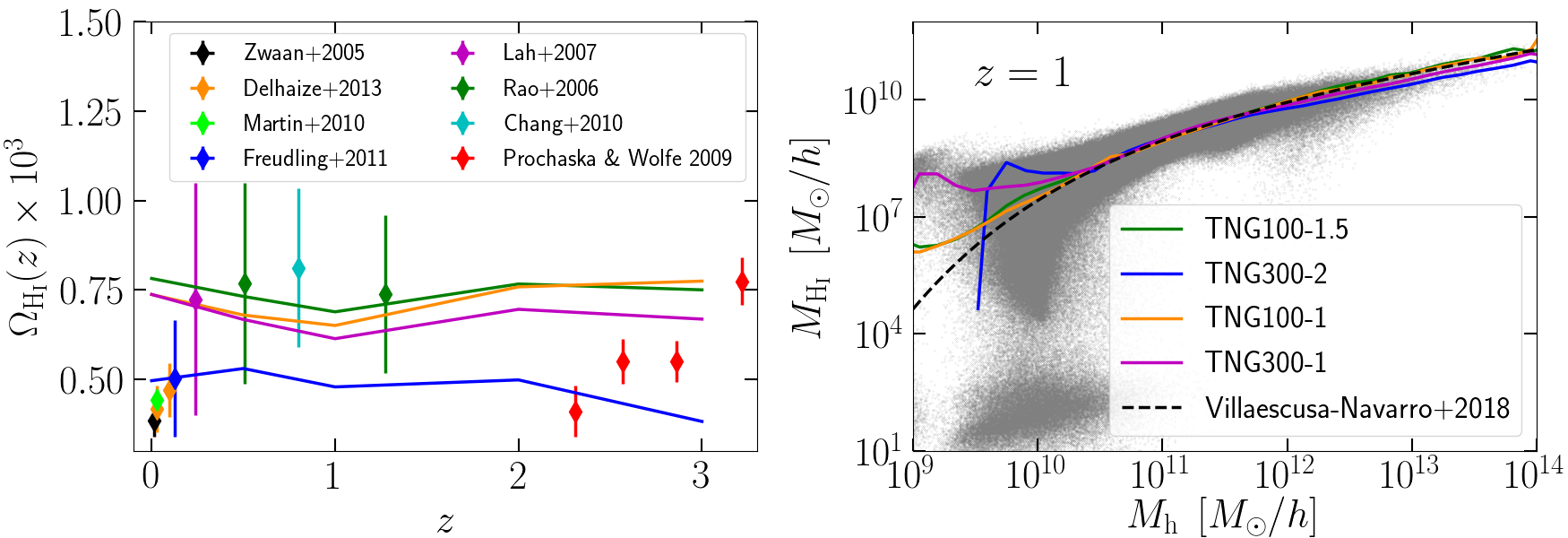}
\caption{Time evolution of the $\hi$ cosmic energy density $\Omega_{\hi}$ (left) and mean $\hi$-to-total halo mass relation $M_{\hi}(M_{\rm h})$ at $z=1$ (right). The result is shown for the fiducial cosmology and for the TNG300-2 (blue), TNG100-1.5 (green), TNG300-1 (purple) and TNG100-1 (orange) resolutions. On the left, the symbols with error bars show a few representative observational estimates compiled in Table 2 of Ref.~\cite{2013MNRAS.433.1398D}, and converted to our fiducial cosmology: black from Ref.~\cite{2005MNRAS.359L..30Z}, orange from Ref.~\cite{2013MNRAS.433.1398D}, light green from Ref.~\cite{2010ApJ...723.1359M}, blue from Ref.~\cite{2011ApJ...727...40F}, purple from Ref.~\cite{2007MNRAS.376.1357L}, green from Ref.~\cite{2006ApJ...636..610R}, cyan from Ref.~\cite{2010Natur.466..463C} and red from Ref.~\cite{2009ApJ...696.1543P} (see Ref.~\cite{2019MNRAS.487.1529D} for a discussion of the comparison between the simulation and these observational results). On the right, the grey dots mark the actual $\hi$ and total mass of the halos in the TNG300-1 simulation, and the black dashed line shows the three-parameter fit obtained in Ref.~\cite{2018ApJ...866..135V} using their modeling of $\hi$ in TNG100-1.}
\label{fig:validation}
\end{figure}

The left panel of Fig.~\ref{fig:validation} shows the time evolution of the total $\hi$ energy density $\Omega_{\hi}(z) = \rho_{\hi}(z)/\rho_{c0}$, where $\rho_{c0} = 3H_0^2/(8\pi G)$ is the critical energy density today. The result is shown for the fiducial cosmology of our TNG300-2 and TNG100-1.5 resolutions (blue and green), as well as for the original TNG300-1 and TNG100-1 simulations (purple and orange). For the latter two cases, our results agree very well with those obtained using the same simulation data in Refs.~\cite{2018ApJ...866..135V, 2019MNRAS.487.1529D}. Recall, in particular, that the modeling in Ref.~\cite{2018ApJ...866..135V} differs from ours in that they recompute $f_{\rm H_N}$ in star-forming gas at post-processing, whereas we simply set it to unity: the good agreement between the two works thus illustrates the weak impact of this approximation. We have also explicitly checked that the $\hi$ content of halos in our simulations agrees very well with that shown in Ref.~\cite{2018ApJ...866..135V}. As an example, we show in the right panel of Fig.~\ref{fig:validation}, the mean $\hi$-to-total halo mass relation $M_{\hi}(M_{\rm h})$ at $z=1$. The black dashed curve shows the fit of Ref.~\cite{2018ApJ...866..135V} to their relation in the TNG100-1 simulation, which agrees very well with our result for the same resolution (orange line) for $M_{\rm h} \gtrsim 10^{10}\ M_{\odot}/h$; the difference at lower halo masses is expected as the simple fit does not describe the details of the low-mass end of the distribution also in Ref.~\cite{2018ApJ...866..135V}.

Figure \ref{fig:validation} shows also that numerical resolution has a strong impact on the $\hi$ distribution. Generically, $\Omega_{\hi}$ decreases with decreasing resolution, as best seen by the markedly lower $\hi$ abundance in TNG300-2 compared to all other resolutions. Note however that TNG100-1.5 is a resolution intermediate to TNG100-1 and TNG300-1, but it has more $\hi$ at $z<1$. The impact of numerical resolution can be understood at least partly by inspecting the behavior of the $M_{\hi}(M_{\rm h})$ relation. For example, the right panel of Fig.~\ref{fig:validation} shows that this relation in TNG300-2 cuts off sharply for $M_{\rm h} \lesssim 5\times 10^{9}\ M_{\odot}/h$. This is because at this resolution, these objects are not as well resolved, and supernovae driven winds, tidal stripping and heating by UV background radiation become very efficient at removing gas, which gets more easily ionized once in the intergalactic medium (see Ref.~\cite{2018ApJ...866..135V} for an indepth discussion of this low-mass cutoff). Additionally, high-density gas clouds that can shield the $\hi$ from ionizing radiation are also harder to resolve overall in TNG300-2. At the higher mass end, starting from $M_{\rm h} \gtrsim 10^{12}\ M_{\odot}/h$, the halos in TNG300-2 also display a lower amount of $\hi$, which is partly due to these larger halos comprising also a larger number of poorly resolved subhalos, which for the reasons just discussed are not as $\hi$ rich. The results from the other numerical resolutions can be broadly explained along similar lines. We shall take these considerations around numerical resolution into account when we interpret our numerical findings below.

\subsection{Bias parameter measurements}
\label{sec:bias}

We evaluate the $\hi$ bias parameters $\blin$, $\bphi$ and $\bphidelta$ in a way that is in all analogous to how Ref.~\cite{2022JCAP...01..033B} evaluated the same parameters for halos and galaxies. Concerning $b_1$, we estimate it as
\bq\label{eq:b1est}
\blin(z) = \lim_{k \to 0} \frac{P_{m\hi}(k,z)}{P_{mm}(k,z)},
\eq
where $P_{m\hi}$ and $P_{mm}$  are the $\hi$-matter cross-power spectrum and matter power spectrum, respectively. The lowest wavenumbers probed by our simulations are $k \gtrsim 0.03 h/{\rm Mpc}$ for TNG300-2 and $k \gtrsim 0.08 h/{\rm Mpc}$ for TNG100-1.5, where the scale-dependence of the $P_{m\hi}/P_{mm}$ ratio can still be non-negligible. To make our estimates of $b_1$ more robust then, we fit this ratio on scales $k < 0.15 h/{\rm Mpc}$ for TNG300-2 and $k < 0.3 h/{\rm Mpc}$ for TNG100-1.5 using $b_1 + A k^2$, where $A$ is a nuisance parameter that absorbs the leading-order impact from scale-dependent corrections.\footnote{We also calculated $b_1$ using the original TNG300-1 simulation (not shown), which is closer in resolution to TNG100-1.5, but has a bigger box size. We found the corresponding values of $b_1$ to agree relatively well, and so that any errors on our $b_1$ estimates deriving from the small size of TNG100-1.5 do not have a significant impact on our final conclusions.} We show as error bars on $b_1$ the error estimated from the least-squares fitting method.

For the case of $\bphi$, we estimate it using its definition as the response of the $\hi$ energy density to changes in the amplitude of the primordial gravitational potential $\phi$ in cosmologies with local PNG (cf.~Eq.~(\ref{eq:respdef})). Concretely, if $\phi_{\rm L}$ is the amplitude of some long-wavelength primordial potential perturbation, then it can be shown using the separate universe ansatz \cite{dalal/etal:2008, slosar/etal:2008} that local structure formation inside this perturbation is equivalent to global structure formation in a cosmology with the primordial scalar power spectrum amplitude $\A_s$ rescalled as $\A_s \to \A_s\left[1 + \delta_{\A_s}\right]$, where $\delta_{\A_s} = 4\fnl\phi_{\rm L}$, i.e., 
\bq\label{eq:bphidef}
\bphi(z) = \frac{\partial{\rm ln}\rho_{\hi}(z)}{\partial(\fnl\phi_L)} \equiv 4\frac{\partial{\rm ln}\rho_{\hi}(z)}{\partial\delta_{\A_s}}.
\eq
The bias parameter $\bphi$ can then be estimated by finite diferencing using our set of fiducial, High$\A_s$ and Low$\A_s$ cosmologies as
\bq\label{eq:bphiest1}
\bphi(z) = \frac{\bphi^{{\rm High}\A_s} + \bphi^{{\rm Low}\A_s}}{2},
\eq
with
\bq
\label{eq:bphiest2} \bphi^{{\rm High}\A_s} &=& \frac{4}{+|\delta_{\A_s}|}\left[\frac{\rho_\hi^{{\rm High}\A_s}(z)}{\rho_\hi^{\rm Fiducial}(z)} - 1\right], \\
\label{eq:bphiest3} \bphi^{{\rm Low}\A_s} &=& \frac{4}{-|\delta_{\A_s}|}\left[\frac{\rho_\hi^{{\rm Low}\A_s}(z)}{\rho_\hi^{\rm Fiducia}(z)} - 1\right],
\eq
and where the superscripts indicate in which cosmology the $\hi$ energy density is measured (recall $|\delta_{\A_s}| = 0.05$). With only one realization of the initial conditions we cannot quote errors on our measurements in a statistical ensemble sense, and bootstrapping or resampling methods are also not adequate approaches given our relatively small box sizes. We note however that $\bphi^{{\rm High}\A_s}$ and $\bphi^{{\rm Low}\A_s}$ should be the same up to numerical noise, and so we will use their difference as a rough estimate of the error in our measurements. Note also that since the three cosmologies share the same phases of the initial conditions, sample variance errors on $\bphi$ cancel already to a large degree.

Finally, we evaluate $\bphidelta$ as
\bq\label{eq:bphideltaest}
\bphidelta(z) &=& \left[\frac{\partial{\rm ln}b_1(z)}{\partial(\fnl\phi)} + \bphi(z) \right] \blin(z) \equiv \left[4\frac{\partial{\rm ln}b_1(z)}{\partial\delta_{\A_s}} + \bphi(z) \right] \blin(z), 
\eq
where the second equality follows again from the separate universe equivalence between long-wavelength $\fnl\phi$ perturbations and changes to $\A_s$. This equation can be derived from the definitions of $b_1$ and $\bphidelta$ in Eq.~(\ref{eq:respdef}) \cite{2022JCAP...01..033B}, noting that $\rho_{\hi} b_{\phi\delta} = \partial\left(\rho_{\hi}b_1\right)/\partial(\fnl\phi)$. We evaluate the derivative of $b_1$ in Eq.~(\ref{eq:bphideltaest}) via finite differences (analogously to Eqs.~(\ref{eq:bphiest1})-(\ref{eq:bphiest3})) using the values of $b_1$ estimated from the fiducial, High${\A_s}$ and Low${\A_s}$ simulations using Eq.~(\ref{eq:b1est}). Note that $b_{\phi\delta}$ could also be estimated with separate universe simulations that account simultaneously for changes to $\A_s$ and the mean cosmic matter density, but at the cost of running additional simulations to the ones we use in this paper.

\section{Results}
\label{sec:results}

In this section we present and discuss our main $\hi$ bias results. We discuss first the parameter $\bphi$ and its relation to $b_1$, and then do the same for the $\bphidelta$ parameter. Our numerical values for these three bias parameters are listed in Tab.~\ref{tab:vals}.

\begin{table*}
\centering
\begin{tabular}{lcccccccccccccc}
\toprule
& {\rm Redshift} & $z = 0$ & $z = 0.5$ & $z = 1$ & $z = 2$ &  $z = 3$  \\
\midrule
& $b_1$ (TNG100-1.5)    & $0.73 \pm 0.07$ & $1.01 \pm 0.04$ & $1.42 \pm 0.02$ & $2.04 \pm 0.02$ &  $2.5 \pm 0.04$  \\
& $b_1$ (TNG300-2) & $0.62 \pm 0.06$ & $0.86 \pm 0.04$ & $1.18 \pm 0.03$ & $1.80 \pm 0.02$ &  $2.58 \pm 0.06$  \\
\midrule
& $\bphi$ (TNG100-1.5) & $-1.70 \pm 0.05$ & $-0.44 \pm 0.39$ & $0.47 \pm 0.12$ & $2.39 \pm 0.12$ &  $3.82 \pm 0.06$  \\
& $\bphi$ (TNG300-2) & $-1.76 \pm 0.16$ & $-1.40 \pm 0.11$ & $-0.26 \pm 0.32$ & $2.08 \pm 0.14$ &  $1.98 \pm 0.22$  \\
\midrule
& $\bphidelta$ (TNG100-1.5) & $-3.86 \pm 0.68$ & $-2.27 \pm 0.47$ & $-1.70 \pm 1.71$ & $0.27 \pm 1.44$ &  $5.37 \pm 1.66$  \\
& $\bphidelta$ (TNG300-2) & $-2.55 \pm 0.65$ & $-3.23 \pm 0.36$ & $-2.93 \pm 0.52$ & $1.06 \pm 0.97$ &  $1.09 \pm 14.5$  \\
\bottomrule
\end{tabular}
\caption{Values of the bias parameters $b_1$, $\bphi$ and $\bphidelta$ of the $\hi$ distribution measured in this work.}
\label{tab:vals}
\end{table*}

\subsection{The $b_\phi$ parameter of $\hi$}
\label{sec:bphi}

The left panel of Fig.~\ref{fig:bphi} shows the redshift evolution of the $\bphi$ values of $\hi$ from the TNG100-1.5 and TNG300-2 resolutions, as labeled. Both resolutions predict that $\bphi$ increases with increasing redshift, and that it is negative at $z \lesssim 1$. The two resolutions are in relatively good agreement in their predictions, although there are visible differences at $z=0.5$, $z=1$ and $z=3$. The right panel of Fig.~\ref{fig:bphi} shows $\bphi$ plotted against the corresponding values of $\blin$, where we note a similar level of agreement between the two resolutions, except perhaps again at $z=3$ (rightmost point) where TNG300-2 displays a markedly lower value of $\bphi$. 

\subsubsection{Halo model interpretation of the results}
\label{sec:bphi_hm}

\begin{figure}
\centering
\includegraphics[width=\textwidth]{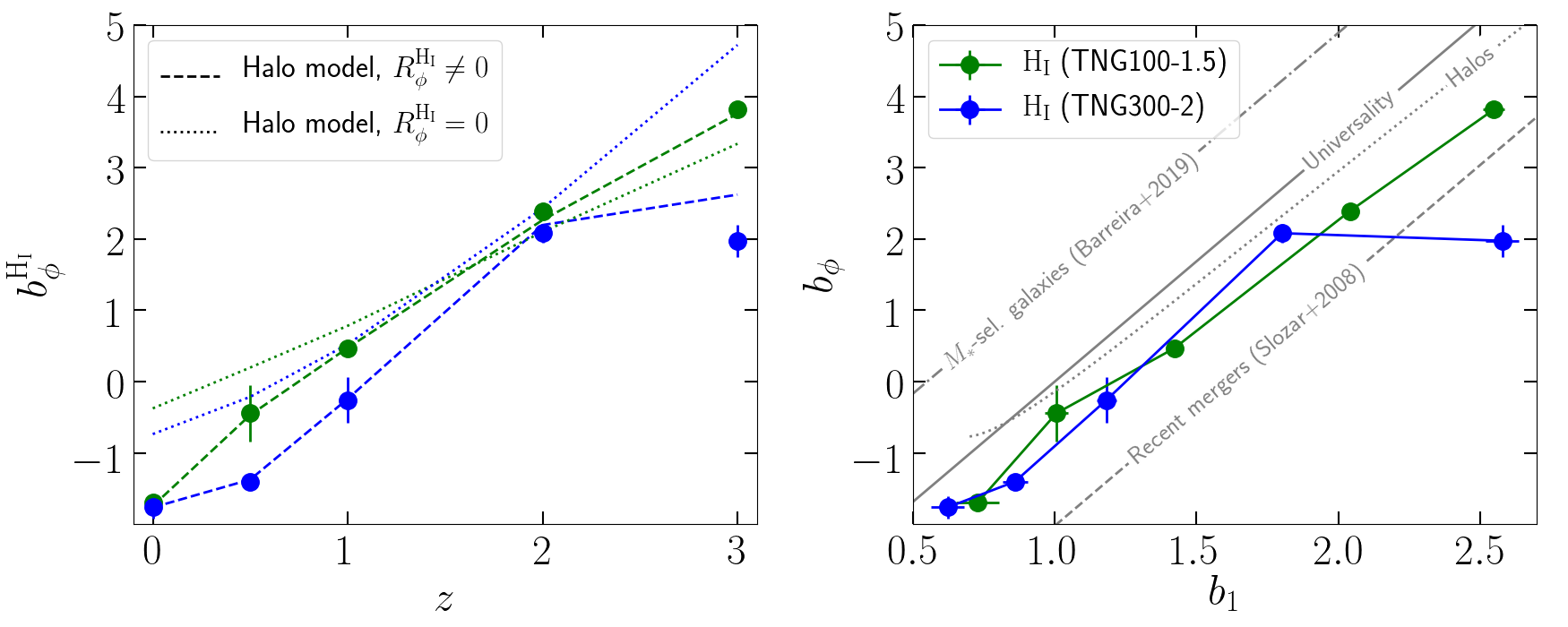}
\caption{Time evolution of the $\bphi$ parameter of $\hi$ (left) and its relation to $b_1$ (right). The result is shown by the filled circles with error bars for the two resolutions TNG100-1.5 (green) and TNG300-2 (blue). On the left panel, the dashed and dotted lines show the predictions of the halo model, respectively, with and without the response $R_{\phi}^{\hi}$ of the $\hi$-to-total halo mass relation taken into account in Eq.~(\ref{eq:bphihm}). On the right, the solid grey line marks the universality relation, the dotted grey line is the expected relation for halos, the grey dashed line shows the recent-merger relation of Ref.~\cite{slosar/etal:2008}, and the relation for stellar-mass selected galaxies in IllustrisTNG from Ref.~\cite{2020JCAP...12..013B} is shown by the grey dot-dashed line.}
\label{fig:bphi}
\end{figure}

The halo model is a useful tool to guide the interpretation of our numerical results. Assuming that all of the $\hi$ is inside halos with some mass $M_{\rm h}$, then $\bphi$ can be written as \cite{Voivodic:2020bec}
\bq\label{eq:bphihm}
\bphi^{\hi}(z) = \frac{\int {\rm d}M_{\rm h} n_{\rm h}(M_h)M_{\hi}(M_{\rm h})\left[\bphi^{\rm h}(M_{\rm h}) + R_{\phi}^{\hi}(M_{\rm h})\right]}{\int {\rm d}M_{\rm h} n(M_h)M_{\hi}(M_{\rm h})},
\eq
where $n_{\rm h}(M_{\rm h})$ is the differential halo mass function, $M_{\hi}(M_{\rm h})$ is the mean $\hi$ mass in halos of mass $M_{\rm h}$ (cf.~Fig.~\ref{fig:validation}), $\bphi^{\rm h}(M_{\rm h})$ is the local PNG bias of the halos and 
\bq\label{eq:Rphi}
R_{\phi}^{\hi}(M_{\rm h}) = \frac{\partial{\rm ln}M_{\hi}(M_{\rm h})}{\partial(\fnl\phi)}
\eq
is the response of the $M_{\hi}(M_{\rm h})$ relation to long-wavelength perturbations $\fnl\phi$. This response function is an ingredient that is often ignored in the literature, but which must be present in general to account for the modulation of $M_{\hi}(M_{\rm h})$ by $\fnl\phi$, in the same way that $\bphi^{\rm h}$ accounts for the same modulation of the halo mass function; see Ref.~\cite{Voivodic:2020bec} for more details. We have evaluated Eq.~(\ref{eq:bphihm}) using its ingredients measured from our TNG100-1.5 and TNG300-2 simulations. In doing so, we considered all halos with at least $N_{\rm cell} = 50$ gas cells, which for TNG100-1.5, accounts for over $99\%$ of all of the $\hi$ at $z<2$ and $96\%$ at $z=3$, and for TNG300-2, it accounts for over $99\%$ at $z<1$, $97\%$ at z=2 and $91\%$ at z=3. The result is shown by the dashed lines in the left panel of Fig.~\ref{fig:bphi}, which agree extremely well with the simulation measurements, as expected.\footnote{The poorer performance for TNG300-2 at $z=3$ has to do with the fact that the halos with $N_{\rm cell} \geq 50$ account only for $91\%$ of all of the $\hi$ in the simulation box, and thus the starting assumption of the halo model breaks down slightly.} 

\begin{figure}
\centering
\includegraphics[width=\textwidth]{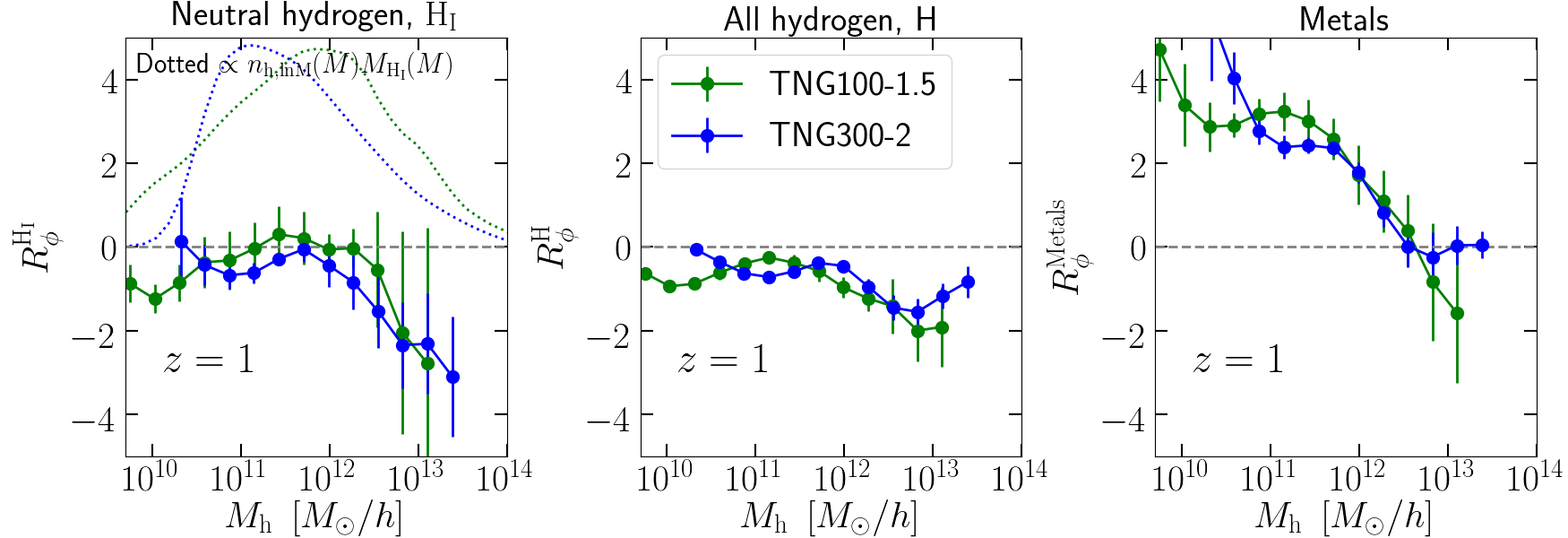}
\caption{Response functions of the halo gas-to-total mass relation to long-wavelength perturbations $\fnl\phi$, $R_{\phi}^{\rm gas} = \partial{\rm ln}M_{\rm gas}(M_{\rm h})/\partial(\fnl\phi)$. The result is shown on the left for atomic neutral hydrogen (${\rm gas} = \hi$), in the middle for all hydrogen (${\rm gas} = {\rm H}$), and on the right for all metals produced by stars (${\rm gas} = {\rm Metals}$). All panels are for $z=1$ and the result is shown for both resolutions TNG100-1.5 and TNG300-2, as labeled. On the left, the dotted lines are  $\propto n_{\rm h, lnM}M_{\hi}$, where $n_{\rm h, lnM}$ is the number of halos per logarithmic mass bin; this helps visualize which halo mass scales contribute the most to Eq.~(\ref{eq:bphihm}).}
\label{fig:Rphi}
\end{figure}

To focus on the impact of the response function $R_{\phi}^{\hi}(M_{\rm h})$, we show as dotted lines on the left panel of Fig.~\ref{fig:bphi} the outcome of Eq.~(\ref{eq:bphihm}), but with the contribution from $R_{\phi}^{\hi}(M_{\rm h})$ set to zero; by comparing to the dashed line, which shows the full result, one can thus directly appreciate the importance of this response function. Concretely:

\begin{enumerate}

\item At $z \leq 1$, the dashed lines are below the dotted lines, which indicates a negative net effect from $R_{\phi}^{\hi}$. That is, the halos become generically $\hi$ poorer inside $\fnl\phi$ perturbations.

\item At $z=2$, the dashed and dotted lines are comparable, which signals a smaller contribution from $R_{\phi}^{\hi}$. That is, the $\hi$ content of halos responds less strongly to $\fnl\phi$ perturbations.

\item At $z=3$, for TNG100-1.5, the net effect from $R_{\phi}^{\hi}$ is positive, i.e., the $\fnl\phi$ perturbations make the halos slightly $\hi$ richer (this is to a smaller extent already visible at $z=2$). For TNG300-2, however, $R_{\phi}^{\hi}$ has a very negative net effect (cf.~dashed blue line below the dotted one).\footnote{This is likely due to the lower resolution of TNG300-2, where in the less well resolved low-mass halos, the enhanced supernovae feedback and tidal stripping effects caused by the $\fnl\phi$ perturbation (i.e.~an increase in $\A_s$) may be unrealistically too efficient at ejecting gas out to lower density regions more exposed to the ionizing radiation.}

\end{enumerate}
We do not show all of the response function measurements for brevity, and also because they are still relatively noisy and more simulations are needed to aid a more detailed inspection of their mass- and redshift-dependence. Just as a single illustrative case, however, the left panel of Fig.~\ref{fig:Rphi} shows the response function $R_{\phi}^{\hi}$ measured using our separate universe simulations at $z=1$. Indeed, and as anticipated in point 1 above, the values of $R_{\phi}^{\hi}$ are either negative or compatible with zero on the mass range that contributes sizeably to the integral of Eq.~(\ref{eq:bphihm}). 

\subsubsection{The connection to the total hydrogen distribution}
\label{sec:bphi_exp}

To understand further these results, we investigated also the response of the total hydrogen content of the halos, i.e., $R_{\phi}^{\rm H} = \partial{\rm ln}M_{\rm H}(M_{\rm h})/\partial(\fnl\phi)$, where $M_{\rm H}(M_{\rm h})$ is the mean mass in hydrogen (not just $\hi$) in halos of mass $M_{\rm h}$. This is shown in the middle panel of Fig.~\ref{fig:Rphi} for $z=1$, where we observe that it has a shape and size that is similar to the response of the $\hi$ content on the left. That is, the suppression of the $\hi$ content in halos inside $\fnl\phi$ perturbations at $z=1$ follows to some degree the suppression of the total amount of hydrogen ${\rm H}$ in those halos. Physically, this behavior of $R_{\phi}^{\rm H}$ is expected for at least two reasons, both associated with the enhanced structure formation inside $\fnl\phi$ perturbations (or equivalently, given an increase in $\A_s$):

\begin{enumerate}

\item First, as found in Ref.~\cite{2020JCAP...12..013B}, these perturbations enhance star formation, which accelerates the transformation of the primordial hydrogen into stars and heavier metals. To help visualize this effect, the right panel of Fig.~\ref{fig:Rphi} shows the response $R_{\phi}^{\rm Metals} = \partial{\rm ln}M_{\rm Metals}(M_{\rm h})/\partial(\fnl\phi)$, where $M_{\rm Metals}(M_{\rm h})$ is the mean mass of all metals produced by stars inside halos of mass $M_{\rm h}$. Indeed, the positive value of $R_{\phi}^{\rm Metals}$ confirms that $\fnl\phi$ perturbations boost the conversion of hydrogen onto heavier metals through star formation.\footnote{This result finds also interesting ramifications to attempts to constrain cosmology and local PNG using intensity mapping observations from the emission lines of some of these heavier elements \cite{2017MNRAS.464.1948F, 2019ApJ...872..126M, 2021arXiv211103717M}.} 

\item The second reason has to do with the fact that these perturbations also enhance the effects of supernovae (for $M_{\rm h} \lesssim 10^{12} M_{\odot}/h$) and black hole feedback (for $M_{\rm h} \gtrsim 10^{12} M_{\odot}/h$) that expel some of the gas out of the halos.

\end{enumerate}

\begin{figure}
\centering
\includegraphics[width=\textwidth]{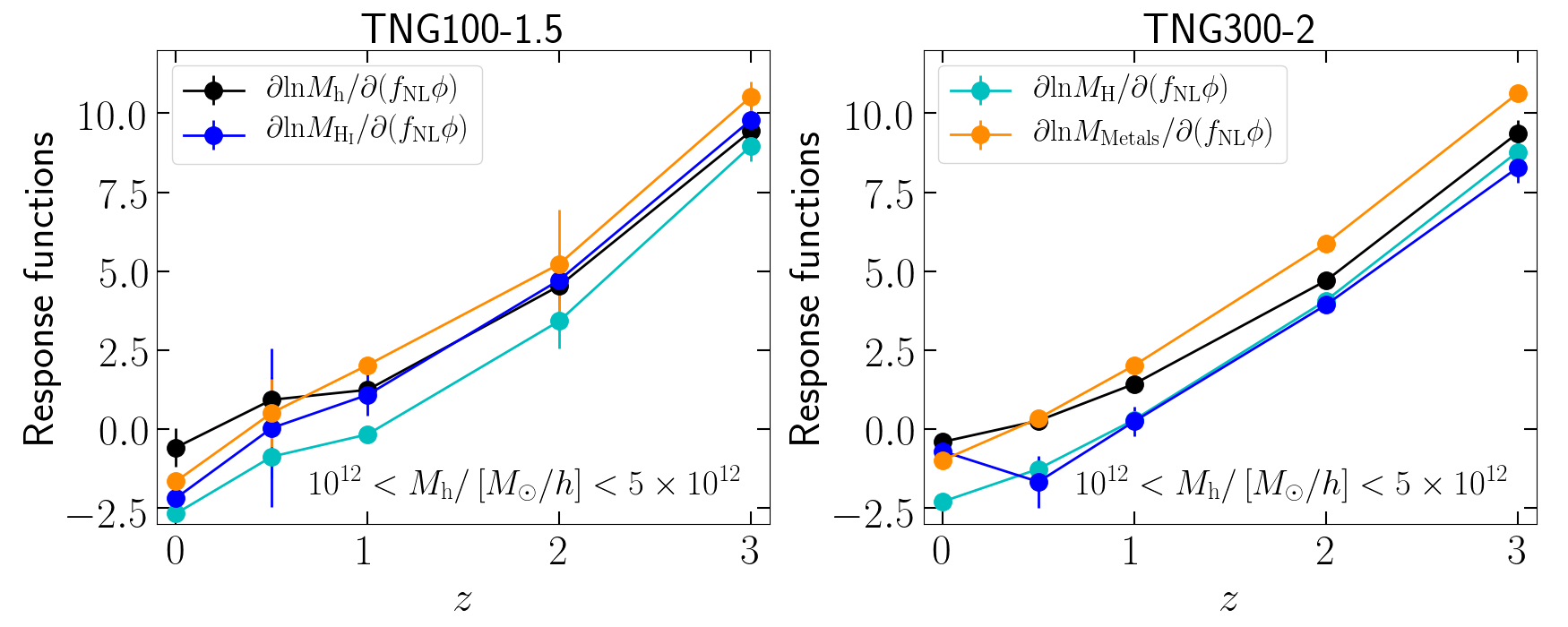}
\caption{Response of the total mass (black), mass in $\hi$ (blue), mass in all hydrogen (cyan) and mass in metals (orange) found inside all halos in the total mass bin $M_{\rm h} \in \left[10^{12}, 5\times 10^{12} \right]M_{\odot}/h$. The result is shown as a function of redshift, and on the left and right for TNG100-1.5 and TNG300-2, respectively.}
\label{fig:Rphi_timeevo}
\end{figure}

For another perspective into these results, we show in Fig.~\ref{fig:Rphi_timeevo} the redshift evolution of the response of the total mass, mass in $\hi$, mass in all hydrogen and mass in metals, found in all halos in the mass bin $M_{\rm h} \in \left[10^{12}, 5\times 10^{12} \right]M_{\odot}/h$, as labeled. Focusing first on the TNG300-2 results on the right, the black line shows that the $\fnl\phi$ perturbations work to increase the total mass (except at $z=0$ when it decreases just slightly), and do more so at higher redshift; this is a consequence of there being more halos in the bin, as well as each halo becoming also more massive. The figure shows also that while the mass in $\hi$ can increase for $z \gtrsim 1$ (cf.~blue line), this increase is always smaller than that in total mass, which effectively causes a suppression of the $\hi$-to-total mass relation, i.e., $R_\phi^{\hi} < 0$ (cf.~Fig.~\ref{fig:Rphi}). At this resolution, the behavior of the $\hi$ response is well traced by that of the total hydrogen shown in cyan color. Note also how the effects of enhanced star formation are visible at $z \geq 1$ by the enhanced mass in metals (cf.~orange line), but at lower redshift the stronger effects by black hole feedback inside $\fnl\phi$ perturbations eventually work to remove metals from the halos, and can lead to the response of the mass in metals to drop below that of the total mass at $z=0$.

The results from the TNG100-1.5 resolution on the left of Fig.~\ref{fig:Rphi_timeevo} agree relatively well with those from TNG300-2, with the main noteworthy difference being that in TNG100-1.5 the behavior of the $\hi$ response is not as well traced by the total hydrogen response. Concretely, at $z \leq 1$ the response of the $\hi$ mass is below that of the total mass, but still visibly above the response of the total hydrogen, i.e., the $\fnl\phi$ perturbations still lower the $\hi$-to-total mass relation of these halos, but lower the H-to-total mass relation even more. Further, at $z \geq 2$ the $\fnl\phi$ perturbations can actually lead to an increase in $\hi$ mass that slightly outweighs the increase in total mass. This could be associated with the fact that $\fnl\phi$ perturbations promote the formation of denser gas clouds that can shield the hydrogen inside them more efficiently from the ionizing radiation. These dense structures are not as well resolved in TNG300-2, which can explain why the same effect is not visible at this lower resolution.

There are more physical effects that can drive differences between the impact that $\fnl\phi$ perturbations have on the $\hi$ and hydrogen mass of halos. For example, related to point 2 above, for the case of the $\hi$ it is actually simply sufficient that the enhanced feedback or tidal stripping effects remove gas from inside high-density regions in the halo (and not completely from inside the halo altogether) for it to be more exposed to the ionizing radiation; this lowers $R_{\phi}^{\hi}$, but not $R_{\phi}^{\rm H}$. Further, $\fnl\phi$ perturbations also modify the details of the evolution of AGN, and thus their contribution to the local ionizing radiation field in IllustrisTNG (which impacts the $\hi$ directly, but not the total hydrogen). In fact, here one should note that since stars form earlier inside $\fnl\phi$ perturbations, reionization begins earlier as well. This effect is not accounted for in our results as we assumed the time-dependent UV background (the main driver of reionization in IllustrisTNG) to be the same in our fiducial, High$\A_s$ and Low$\A_s$ cosmologies. This should not have a critical impact in our results as we focus on redshifts well after reionization is complete, but it is interesting to investigate the impact of this approximation in works more focused around the epoch of reionization $6 \lesssim z \lesssim 10$. We defer to future work a more indepth investigation of the complicated interplay between these and other astrophysical effects, which we note will depend in general on the assumed galaxy formation model and $\hi$ modeling strategy.

\subsubsection{Comparison to the universality relation}
\label{sec:bphi_univ}

Finally, we compare the $\bphi(\blin)$ relation of $\hi$ with other relations in the literature. The solid grey line in the right panel of Fig.~\ref{fig:bphi} shows the popular universality relation $\bphi = 2\delta_c\left(\blin - 1\right)$, where $\delta_c = 1.676$ is the critical density for spherical collapse. The typical approach in the literature (see e.g.~Refs.~\cite{2019PhRvD.100l3522B, 2021PhRvD.103f3520L, 2020MNRAS.496.4115C}) estimates the $\bphi(b_1)$ relation of the $\hi$ using Eq.~(\ref{eq:bphihm}) with $R_{\phi}^{\hi} = 0$ and assuming the universality relation for the halo bias $b_{\rm h}$. Since the universality relation is linear, is follows trivially then that $\bphi(b_1)$ of the $\hi$ is given by the exact same relation, irrespective of the form of $M_{\hi}(M_{\rm h})$. This can be checked by assuming $b_\phi^{\rm h}(M_{\rm h}) = A + Bb_1^{\rm h}(M_{\rm h})$ for the bias of the halos in Eq.~(\ref{eq:bphihm}) with $R_\phi^{\hi} = 0$, and see that this always results in the same relation $b_\phi^{\hi} = A + Bb_1^{\hi}$ for the $\hi$.

The figure shows that this calculation overpredicts the amplitude of the $\hi$ $\bphi(b_1)$ relation measured from our IllustrisTNG simulations, which we checked (not shown) is roughly bracketed by $\bphi = 2\delta_c\left(b_1 - p\right)$ with $p \in \left[1.1, 1.4\right]$. For $b_1 \lesssim 1.5$ ($z \lesssim 1$), this is largely because of the negative values of $R_\phi^{{\hi}}$, which push the $\bphi(b_1)$ relation downwards. For $b_1 \gtrsim 1.5$ ($z\gtrsim1$) the result is a combination of the impact of $R_\phi^{\hi}$ and the relation of the halos shown by the grey dotted line\footnote{To evaluate the $\bphi(b_1)$ relation of the halos, we calculate $\bphi = 4\partial{\rm ln}n^{\rm Tinker}(M_{\rm h})/\partial{\delta_{\A_s}}$, where $n^{\rm Tinker}$ is the halo mass function of Ref.~\cite{2008ApJ...688..709T}, and obtain $b_1$ using the fitting formula of Ref.~\cite{2010ApJ...724..878T}.}, which is itself already below the universality relation. Note also that the predictions for the bias parameter $b_1$ also affect the $b_\phi(b_1)$ relation. For comparison purposes only, the right panel of Fig.~\ref{fig:bphi} shows also the {\it recent-merger} relation $\bphi = 2\delta_c\left(1 - 1.6\right)$ (dashed) derived by Ref.~\cite{slosar/etal:2008} and that is usually assumed in constraints using quasars \cite{2021arXiv210613725M, 2019JCAP...09..010C}, and the relation $\bphi = 2\delta_c\left(1 - 0.55\right)$ (dot-dashed), which was found in Ref.~\cite{2020JCAP...12..013B} to roughly describe the case for stellar-mass selected galaxies in IllustrisTNG.

\subsection{The $b_{\phi\delta}$ parameter of $\hi$}
\label{sec:bphi}

\begin{figure}
\centering
\includegraphics[width=\textwidth]{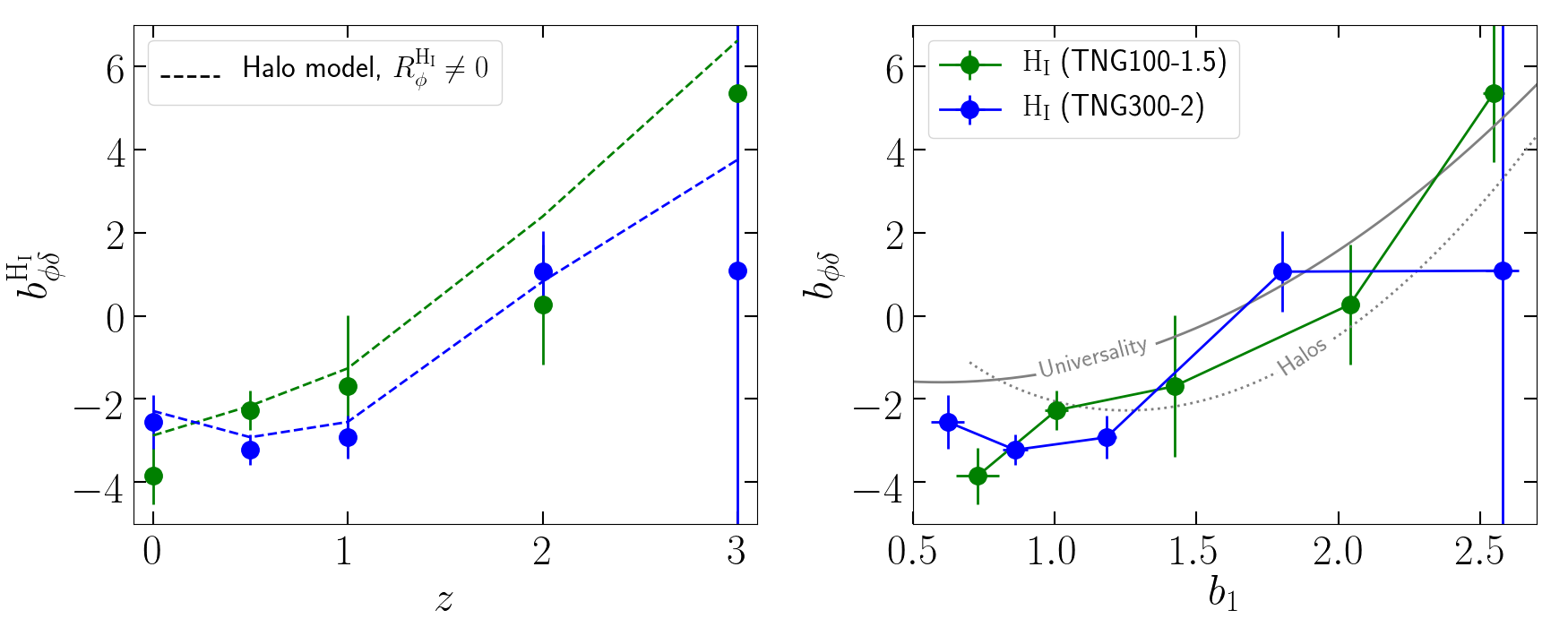}
\caption{Same as Fig.~\ref{fig:bphi}, but for $\bphidelta$ instead of $\bphi$. On the left, the dashed lines show the halo model prediction of Eq.~(\ref{eq:bphideltahm}) with $R_{\phi}^{\hi}$ given by the simulation measurements, but with $R_{1}^{\hi} = R_{\phi\delta}^{\hi}=0$. On the right, the solid grey line marks the universality relation, and the grey dotted line shows the expected relation for halos obtained with the definition of Eq.~(\ref{eq:bphideltaest}) using the fitting formulae from Refs.~\cite{2008ApJ...688..709T, 2010ApJ...724..878T}.}
\label{fig:bphidelta}
\end{figure}

Our results for the $\bphidelta$ parameter of $\hi$ are displayed in Fig.~\ref{fig:bphidelta}, which has the same format as Fig.~\ref{fig:bphi} discussed above for $\bphi$. The first noteworthy point is that our $\bphidelta$ estimates are noisier compared to $\bphi$, which is expected since $\bphidelta$ is a second-order bias parameter; this is especially dramatic in our results for TNG300-2 at $z=3$. Despite the noise, there are still a few trends that one can discern from our results, in particular, that $\bphidelta$ is generically a growing function of redshift ($\bphidelta < 0$ at $z \lesssim 2$), and that for $b_1 \lesssim 1.5$, the simulation measurements predict a $\bphidelta(b_1)$ relation that falls below the corresponding universality expression (grey solid line in Fig.~\ref{fig:bphidelta}). The latter is in this case given by $\bphidelta(b_1) = \bphi - b_1 + 1 + \delta_c[b_2(b_1) - (8/21)(b_1-1)]$, where $b_2$ is the bias parameter associated second-order mass perturbations $\delta_m(\vx, z)^2$, and which we evaluate using the polynomial fit $b_2(b_1) = 0.412 - 2.143b_1 + 0.929b_1^2 + 0.008b_1^3$ of Ref.~\cite{lazeyras/etal}. This fit was obtained for dark matter halos using separate universe simulations that do not assume universality of the halo mass function, but the impact of this is small for the purpose of our comparisons here (see Ref.~\cite{lazeyras/etal} for a discussion).

Similarly to as discussed in the last section for $\bphi$, here too the halo model is a useful tool to interpret our numerical results. The halo model expression for $\bphidelta$ is given by \cite{Voivodic:2020bec}
\bq\label{eq:bphideltahm}
\bphidelta^{\hi}(z) = \frac{\int {\rm d}M_{\rm h} n_{\rm h}(M_h)M_{\hi}(M_{\rm h})\left[\bphidelta^{\rm h}(M_{\rm h}) + b_1^{\rm h}(M_{\rm h})R_{\phi}^\hi(M_{\rm h}) + b_\phi^{\rm h}(M_{\rm h})R_{1}^\hi(M_{\rm h}) + R_{\phi\delta}^{\hi}(M_{\rm h})\right]}{\int {\rm d}M_{\rm h} n(M_h)M_{\hi}(M_{\rm h})}, \nonumber \\
\eq
where $b_1^{\rm h}$, $\bphi^{\rm h}$ and $\bphidelta^{\rm h}$ are the bias parameters of the halos, and in addition to $R_{\phi}^\hi$ defined in Eq.~(\ref{eq:Rphi}), now the result depends also on two extra response functions of the $\hi$-to-total halo mass relation defined as
\bq
\label{eq:R1}R_{1}^{\hi}(M_{\rm h}) &=& \frac{\partial{\rm ln}M_{\hi}(M_{\rm h})}{\partial\delta_m},\\
\label{eq:Rphidelta} R_{\phi\delta}^{\hi}(M_{\rm h}) &=& \frac{1}{M_{\hi}(M_{\rm h})}\frac{\partial^2M_{\hi}(M_{\rm h})}{\partial\delta_m\partial(\fnl\phi)}.
\eq
The set of separate universe simulations we use in this work does not let us estimate these response functions reliably, which keeps us from being able to evaluate Eq.~(\ref{eq:bphideltahm}) entirely. We evaluate it nonetheless with $R_{1}^{\hi} = R_{\phi\delta}^{\hi}=0$ and the $R_\phi^{\hi}$ measured from the simulations, and the result is shown by the dashed lines in the left panel of Fig.~\ref{fig:bphidelta}.\footnote{To perform this calculation we evaluate $b_1^{\rm h}$ using the fitting formulae of Ref.~\cite{2010ApJ...724..878T} and take $\bphidelta^{\rm h}$ to be the dotted line on the right panel of Fig.~\ref{fig:bphidelta}. All other ingredients in Eq.~(\ref{eq:bphideltahm}) are directly measured from the simulations.} The agreement between this halo model prediction and the simulation results is not perfect, but that is not surprising since the halo model result is incomplete; the difference between the dashed lines and the simulation measurements can in fact be used to estimate roughly the size of the combined contribution from $b_\phi^{\rm h}R_{1}^\hi + R_{\phi\delta}^{\hi}$ that was neglected.

Concerning the comparison to the universality prediction, previous works in the literature \cite{2020JCAP...11..052K, 2021PDU....3200821K} have estimated the value of $\bphidelta$ for the $\hi$ using Eq.~(\ref{eq:bphideltahm}) with $R_{\phi}^{\hi} = R_{1}^{\hi} = R_{\phi\delta}^{\hi}=0$ and assuming the universality expression for the bias of the halos $\bphidelta^{\rm h}$. Since the second-derivative of the universality relation w.r.t.~$b_1$ is positive, this pushes the corresponding relation of the $\hi$ upwards. This can be seen by assuming $b_\phi^{\rm h}(M_{\rm h}) = A + Bb_1^{\rm h}(M_{\rm h}) + C\left[b_1^{\rm h}(M_{\rm h})\right]^2$ for the bias of the halos in Eq.~(\ref{eq:bphideltahm}), and noting that if $C > 0$, then at fixed numerical value of $b_1$ one has $\bphidelta^{\hi} > \bphidelta^{\rm h}$. The right panel of Fig.~\ref{fig:bphidelta} shows that for $b_1 \lesssim 2$ ($z \lesssim 2$) our simulation results begin to drop below the universality prediction, and so that this way of calculating the $\bphidelta$ parameter in the literature overestimates (the result becomes less negative) the $\bphidelta(b_1)$ relation of the $\hi$ in our simulations. We leave a more indepth discussion of the $\bphidelta$ parameter of the $\hi$ distribution to future work, which would benefit from larger simulation volumes for improved statistics.

\section{Summary and conclusions}
\label{sec:conc}

Upcoming 21cm line-intensity mapping observations will determine the large-scale distribution of atomic neutral hydrogen $\hi$, which is a very sensitive probe of the local PNG parameter $\fnl$. These future data have the potential to tighten up the current best constraint on $\fnl$ using CMB data, and consequently, to improve on our knowledge of the physics of the early Universe and inflation. The main observational signatures appear in the power spectrum and bispectrum of the $\hi$ distribution and are proportional to terms like $b_1\bphi\fnl$ and $b_1^2\bphidelta\fnl$, where $b_1$, $\bphi$ and $\bphidelta$ are three bias parameters that describe how the $\hi$ energy density depends on large-scale mass overdensities $\delta_m$ and gravitational potential perturbations $\fnl\phi$ (cf.~Eqs.~(\ref{eq:biasexp}) and (\ref{eq:respdef})). Naturally, given the strong degeneracy with $\fnl$, accurate and precise theoretical priors on the expected values of these parameters are of the utmost importance to infer the numerical value of $\fnl$ from the data. Our main goal in this paper was to take the first steps towards estimating these bias parameters using hydrodynamical simulations of cosmic structure formation.

Concretely, in this paper we coupled separate universe simulations of the IllustrisTNG galaxy formation model with the analytical KMT model of the split of neutral hydrogen into atomic ($\hi$) and molecular (${\rm H_2}$) (cf.~Sec.~\ref{sec:hi}) to predict the $\bphi(b_1)$ and $\bphidelta(b_1)$ bias parameter relations of the $\hi$. We used simulation data obtained at two numerical resolutions called TNG300-2 ($L_{\rm box} = 205\ {\rm Mpc}/h$, $N_{\rm p} = 2\times 1250^3$) and TNG100-1.5 ($L_{\rm box} = 75\ {\rm Mpc}/h$, $N_{\rm p} = 2\times 1250^3$), and discussed our numerical findings with the aid of the halo model by inspecting the behavior of the $\hi$ content of dark matter halos. Our main results can summarized as follows:

\begin{itemize}

\item The values of $\bphi$ and $\bphidelta$ grow with redshift and are negative at $z \lesssim 1$ and $z \lesssim 2$, respectively (cf.~Tab.~\ref{tab:vals}, and Figs.~\ref{fig:bphi} and \ref{fig:bphidelta}). Our two numerical resolutions agree also relatively well, despite some visible quantitative differences. 

\item The popular universality expressions for the $\bphi(b_1)$ and $\bphidelta(b_1)$ relations overpredict our simulation measurements (cf.~Figs.~\ref{fig:bphi} and \ref{fig:bphidelta}), indicating that the most common calculation found in the literature may currently overestimate the values of $\bphi$ and $\bphidelta$.

\item The lower values w.r.t.~the universality expressions are in part due to the negative response $R_{\phi}^{\hi}$ of $M_{\hi}(M_{\rm h})$ to $\fnl\phi$ perturbations at $z \lesssim 1$ (cf.~Fig.~\ref{fig:Rphi}), i.e., halos become generically $\hi$ poorer inside these perturbations. We discussed how this can be explained by the fact that these perturbations accelerate the conversion of hydrogen to heavier elements by star formation, and enhance the effects of feedback that ejects the gas out to regions where it is more easily ionized.

\end{itemize}
We note that although the impact of astrophysical uncertainties on $\fnl$ constraints with line-intensity mapping data has been investigated in past works \cite{2020MNRAS.496.4115C, 2019PhRvD.100l3522B, 2021PhRvD.103f3520L}, this was done only through parametrizations of the gas-to-halo connection, i.e.~$M_{\hi}(M_{\rm h})$ for the case of $\hi$. Our finding here that response functions like $R_{\phi}^{\hi}$ and $R_{\phi}^{\rm Metals}$ can be generically nonzero introduces the environmental dependence of the gas-to-halo connection as an additional, new astrophysical ingredient whose impact on $\fnl$ constraints needs to be understood as well. This can be straightforwardly achieved already with the aid of Eqs.~(\ref{eq:bphihm}) and (\ref{eq:bphideltahm}), together with parametrizations for response functions like $R_{\phi}^{\hi}$ \cite{Voivodic:2020bec}. 

\vspace{2mm}

Taken at face value, is our finding that the universality-based calculations overpredict the $\bphi(b_1)$ and $\bphidelta(b_1)$ relations in our simulations good or bad news for $\fnl$ constraints using $\hi$ data? This is hard to answer generically since it depends on the survey setup and redshift range considered. For example, focusing just on the case of $\bphi(b_1)$ for power spectrum analyses (cf.~Fig.~\ref{fig:bphi}), for $b_1 < 1$, which corresponds roughly to $z \lesssim 0.5-1$, the universality relation is negative, and since our $\bphi$ measurements are below it, they are larger in absolute value. This means the $\hi$ distribution is actually more sensitive to the effects of $\fnl$. Conversely, however, starting from $b_1 > 1$, which corresponds roughly to $z \gtrsim 0.5-1$, the universality relation becomes positive, and so by assuming it, one overestimates the values of $\bphi$, and consequently, the true constraining power of the data on $\fnl$. One may argue that most of this constraining power comes from high redshift, where one can access the large distance scales where $\fnl$ contributes most sizeably, but these are also the scales that are expected to be more affected by foreground subtraction systematics. The takeaway is that there is strong motivation to revisit the impact of $\bphi(b_1)$ assumptions in existing forecast studies of $\fnl$ using $\hi$ data. Doing so will also serve the purpose to quantify the accuracy and precision with which future simulation-based works need to determine the local PNG $\hi$ bias parameters in order for line-intensity mapping surveys to meet a certain target constraining power on $\fnl$ (see Refs.~\cite{2020JCAP...12..031B, 2022JCAP...01..033B} for discussions).

\vspace{2mm}

We emphasize that the modeling of the $\hi$ in cosmological hydrodynamical simulations is a field that is still under development and therefore with still many associated uncertainties. The results and discussion in this paper provide already very useful intuition and guidance for future works on the subject, but having been obtained at fixed galaxy formation model and $\hi$ modeling strategy, they should be regarded as just the first step towards more robust theoretical predictions for the $\bphi$ and $\bphidelta$ parameters of the $\hi$. The importance of these bias parameters in observational searches of local PNG strongly motivates giving continuation to the work started here, including with investigations of the sensitivity of the bias parameters to the assumed galaxy formation physics, and the exploration of more sophisticated $\hi$ modeling strategies.

\acknowledgments
We would like to thank Elisa Ferreira, Eiichiro Komatsu and Francisco Villaescusa-Navarro for useful comments and conversations. The author acknowledges support from the Excellence Cluster ORIGINS which is funded by the Deutsche Forschungsgemeinschaft (DFG, German Research Foundation) under Germany's Excellence Strategy - EXC-2094-390783311. The numerical analysis of the simulation data presented in this work was done on the Cobra supercomputer at the Max Planck Computing and Data Facility (MPCDF) in Garching near Munich.

\bibliographystyle{utphys}
\bibliography{REFS}

\end{document}